\documentclass[aps,prd,onecolumn,superscriptaddress]{revtex4}
\usepackage{graphicx}
\usepackage{dcolumn}
\usepackage{bm}
\usepackage{color}
\usepackage{multirow}
\usepackage[normalem]{ulem} 
\usepackage[dvipsnames]{xcolor} 
\usepackage{hyperref}
\hypersetup{
  colorlinks=true,        
  linkcolor=blue,         
  citecolor=cyan,         
}
\usepackage{mathrsfs}  
\usepackage[utf8]{inputenc}
\usepackage{mathtools}
\usepackage{doi}
\usepackage{enumitem}
\usepackage{amsmath}
\usepackage{amssymb}

\begin{document}
\title{Proper-Time Approach in Asymptotic Safety via Black Hole Quasinormal Modes and Grey-body Factors}

\author{Bekir Can Lütfüoğlu}
\email{bekir.lutfuoglu@uhk.cz (Corresponding author)}
\affiliation{Department of Physics, Faculty of Science, University of Hradec Kralove, Rokitanskeho 62/26, 500 03 Hradec Kralove, Czech Republic}

\author{Erdinç Ulaş Saka}
\email{ulassaka@istanbul.edu.tr}
\affiliation{Department of Physics, Faculty of Science, Istanbul University, Vezneciler, 34134 Istanbul, Türkiye}

\author{Abubakir Shermatov}
\email{shermatov.abubakir98@gmail.com}
\affiliation{Institute of Fundamental and Applied Research, National Research University TIIAME, Kori Niyoziy 39, Tashkent 100000, Uzbekistan}
\affiliation{University of Tashkent for Applied Sciences, Str. Gavhar 1, Tashkent 100149, Uzbekistan}
\affiliation{Tashkent State Technical University, Tashkent 100095, Uzbekistan}

\author{Javlon Rayimbaev}
\email{javlon@astrin.uz}
\affiliation{
Urgench State University, Kh. Alimjan Str. 14, Urgench 221100, Uzbekistan}


\author{Inomjon Ibragimov}
\email{i.ibragimov@kiut.uz}
\affiliation{Kimyo International University in Tashkent, Shota Rustaveli street 156, Tashkent 100121, Uzbekistan}

\author{Sokhibjan Muminov} \email{sokhibjan.muminov@mamunedu.uz}   \affiliation{Mamun University, Bolkhovuz street 2, Khiva 220900, Uzbekistan } 


\begin{abstract}
We study the quasinormal mode spectrum and grey-body factors of black holes in an effectively quantum-corrected spacetime, focusing on the influence of near-horizon modifications on observable quantities. Employing scalar, electromagnetic, and Dirac test fields, we analyze the perturbation equations and extract the fundamental quasinormal frequencies using both the 6th-order WKB method with Padé resummation and time-domain integration. Our results show that quantum corrections near the horizon significantly affect the real and imaginary parts of the quasinormal modes, particularly for low multipole numbers and in the near-extremal regime. We also verify the robustness of the correspondence between quasinormal modes and grey-body factors by comparing WKB results with those reconstructed from the dominant quasinormal modes. Across all field types and parameter ranges considered, the WKB method proves accurate within a few percent, confirming its reliability in probing the impact of near-horizon physics. These findings support the use of quasinormal ringing and Hawking radiation spectra as sensitive tools for testing quantum modifications of black hole spacetimes.
\end{abstract}

\maketitle

\section{Introduction}
\label{sec:intro}

Quasinormal modes (QNMs) of black holes have emerged as a key observable in gravitational wave astronomy \cite{LIGOScientific:2016aoc,LIGOScientific:2017vwq}, encoding information about the structure and stability of spacetime in the vicinity of compact objects. They represent the damped oscillations that dominate the late-time response of a black hole to external perturbations and are entirely determined by the background geometry and the nature of the perturbing field. In particular, the complex frequencies of these oscillations provide a unique fingerprint of the black hole's mass, spin, and charge, as well as of any deviations from classical general relativity due to quantum corrections or modified gravity theories~\cite{Konoplya:2011qq,Kokkotas:1999bd,Berti:2009kk,Bolokhov:2025uxz}. Consequently, QNMs have been extensively studied for various quantum corrected black holes, including various approaches to constructing quantum-corrected black hole metric in Asymptotically Safe Gravity and various other Effective Quantum theories of gravity,  \textbf{\cite{Lutfuoglu:2025hwh,Skvortsova:2024msa,Stashko:2024wuq,Konoplya:2022hll,Zhu:2024wic,Momennia:2022tug,Luo:2024dxl,Malik:2025dxn,Malik:2024nhy,Konoplya:2025afm,Konoplya:2024lch,Yang:2024ofe,Filho:2024ilq,Konoplya:2020fwg,Lutfuoglu:2025ohb,Konoplya:2023aph,Malik:2024elk,Bolokhov:2025egl,Dubinsky:2025nxv,Skvortsova:2024atk}.}

In a recent work~\cite{Bonanno:2025dry}, a new family of regular black hole solutions was constructed by incorporating quantum gravitational effects via the proper-time renormalization group flow in the framework of asymptotically safe gravity. The resulting spacetimes, characterized by a running Newton's constant and matched across a regular dust-collapse interior and a static exterior geometry, were shown to be free from curvature singularities and to approach the Schwarzschild solution at large distances asymptotically. Axial gravitational perturbations in this background were analyzed in detail, revealing significant deviations in the QNM spectrum as the quantum parameter controlling the flow approaches its critical value. These deviations were also found to suppress the Hawking emission rate and alter observational features such as the shadow radius and ISCO location, potentially relaxing constraints on primordial black holes as dark matter candidates.

While gravitational perturbations offer insights into the dynamics of spacetime itself, the propagation of test fields—such as scalar, electromagnetic, and Dirac fields—on the same background plays an equally important role in probing the geometry and its phenomenological implications. Unlike the gravitational sector, test fields can be treated without invoking an effective stress-energy tensor or assumptions about anisotropy, thereby providing a complementary and often more robust check on the influence of quantum corrections. Furthermore, since Standard Model fields are naturally described by scalar, vector, and spinor equations, their quasinormal spectra are directly relevant for understanding potential observational signatures in high-energy astrophysical processes.

In this work, we extend the previous analysis by studying the QNMs of massless test scalar, electromagnetic, and Dirac fields propagating in the quantum-corrected black hole geometry introduced in~\cite{Bonanno:2025dry}. We derive the corresponding wave equations and effective potentials, identify the appropriate boundary conditions for quasinormal ringing, and compute the spectra using several complementary methods: the WKB approximation with Padé resummation, and time-domain integration with Prony analysis. We investigate how the real and imaginary parts of the frequencies vary with the quantum deformation parameter and compare the behavior across different spin fields. Our goal is to assess the universality and stability of the quantum modifications observed in the gravitational sector and to evaluate their detectability through test field probes.

While QNMs characterize the intrinsic oscillation response of a black hole to perturbations, grey-body factors encode how this response is filtered by the potential barrier surrounding the black hole, thus modifying the observable Hawking radiation. In particular, the transmission probabilities defined by the grey-body factors determine the energy and particle fluxes measured by distant observers. In quantum-corrected spacetimes, even modest deformations of the near-horizon geometry can lead to significant changes in the effective potential and, consequently, in the grey-body factors. Studying these transmission probabilities, therefore, offers a complementary and physically meaningful avenue for testing deviations from classical general relativity. Moreover, verifying the correspondence between grey-body factors and QNMs in this context provides an important consistency check and insight into the robustness of semiclassical descriptions in modified gravity scenarios. Our goal is to assess how reliably the standard methods for computing grey-body factors, particularly the WKB approximation and its correspondence with quasinormal spectra, remain applicable when quantum effects alter the geometry.

The paper is organized as follows: in Sec.~\ref{sec:background}, we briefly review the quantum-corrected black hole background. Sec.~\ref{sec:waveeq} presents the perturbation equations and effective potentials for scalar, electromagnetic, and Dirac fields. In Sec.~\ref{sec:methods}, we describe the numerical methods used for QNM calculations. Sec.~\ref{sec:results} provides the computed spectra and discusses the dependence of the modes on the quantum parameter, while Sec.~\ref{sec:greybody} presents calculations of grey-body factors and tests their correspondence with QNMs. We conclude in Sec.~\ref{sec:conclusion} with a summary and outlook for future work.

\section{Quantum-Corrected Black Hole Background}
\label{sec:background}

The black hole geometry considered in this work arises from an effective renormalization group (RG) improvement of general relativity within the framework of \emph{asymptotic safety}. Originally proposed by Weinberg~\cite{Weinberg:1979}, this scenario assumes that gravity is governed at high energies by a non-Gaussian ultraviolet fixed point, thereby making the theory nonperturbatively renormalizable~\cite{Reuter:1998, Niedermaier:2006}. In particular, the proper-time formulation of the functional renormalization group~\cite{Bonanno:2019ukb} offers a gauge- and parametrization-independent flow equation that has proven effective in constructing regular, quantum-corrected geometries. For a detailed analysis of regulator, gauge, and parametrization dependence of the proper-time flow in quantum Einstein gravity---showing minor sensitivity to the regulator but a stronger impact of linear vs. exponential parametrizations near the non-Gaussian fixed point---see~\cite{Bonanno:2025tfj}.

In the approach developed in~\cite{Bonanno:2025dry}, a position-dependent Newton’s constant $G(\epsilon)$ is obtained by integrating the flow equations in a collapsing dust interior, where $\epsilon$ is the proper energy density. The interior solution is matched across a timelike hypersurface to a static, spherically symmetric exterior metric, leading to a family of regular black holes free from curvature singularities. The resulting line element in the exterior region is given by
\begin{equation}
    ds^2 = -f(r)\,dt^2 + \frac{dr^2}{f(r)} + r^2 d\Omega^2,
\end{equation}
with the lapse function $f(r)$ taking the form
\begin{equation}
    f(r) = \frac{3M^2 + q r^4 - M \sqrt{9M^2 + q^2 r^6}}{q r^4}
    + \frac{2}{3} q r^2 \, \mathrm{arctanh}\left( \frac{(q - \sqrt{q^2 + 9M^2/r^6}) r^3}{3M} \right),
\end{equation}
where $M$ is the ADM mass and $q$ is a quantum deformation parameter related to the asymptotic safety scale.

This geometry reduces to the Schwarzschild solution in the classical limit $q \to \infty$, and remains regular at all $r > 0$. Depending on the value of $q$, the spacetime can possess zero, one, or two horizons. A critical value $q_{\text{cr}} \approx 1.37\,M$ marks the extremal configuration. Below this value, the metric describes a horizonless compact object, while above it, two horizons (inner and outer) appear, as in a typical regular black hole scenario.

The effective geometry encapsulates quantum gravitational corrections in a self-consistent way. It serves as a tractable model for investigating the observational implications of asymptotic safety in the strong-gravity regime. Previous studies have analyzed axial gravitational perturbations in this background~\cite{Bonanno:2025dry}, finding significant deviations in the quasinormal spectra, grey-body factors, and Hawking radiation rates as $q$ approaches $q_{\text{cr}}$.

In the present work, we extend the analysis to test fields of various spins. Since the scalar, electromagnetic, and Dirac fields do not couple directly to the background matter distribution, they offer a robust probe of the underlying geometry and provide valuable insight into the potential signatures of quantum gravity.

\section{Perturbation Equations for Test Fields}
\label{sec:waveeq}

In order to probe the quantum-corrected black hole geometry introduced in the previous section, we study linear perturbations of massless test fields of spin $s = 0, 1, \tfrac{1}{2}$: scalar, electromagnetic, and Dirac fields, respectively. Since these fields do not couple to the effective matter source that gives rise to the background geometry, their evolution is governed by standard equations in curved spacetime. In each case, the radial part of the perturbation reduces to a Schrödinger-like equation with an effective potential that depends on the background metric function $f(r)$.

\subsection{General Formalism}

We consider a general static, spherically symmetric spacetime of the form
\begin{equation}
ds^2 = -f(r) \, dt^2 + \frac{dr^2}{f(r)} + r^2 d\Omega^2,
\label{eq:metric}
\end{equation}
where the specific form of the lapse function $f(r)$ is given in equation~(2) above for the quantum-corrected solution. The tortoise coordinate $r_*$ is defined by
\begin{equation}
\frac{dr_*}{dr} = \frac{1}{f(r)},
\label{eq:tortoise}
\end{equation}
so that the radial wave equations take a unified form for each spin-$s$ field:
\begin{equation}
\frac{d^2 \Psi}{dr_*^2} + \left[\omega^2 - V_s(r)\right] \Psi = 0,
\label{eq:master}
\end{equation}
where $\Psi$ is the radial part of the perturbing field, $\omega$ is the complex quasinormal frequency, and $V_s(r)$ is the corresponding effective potential.

\subsection{Scalar Field ($s=0$)}

The massless scalar field $\Phi$ satisfies the covariant Klein–Gordon equation:
\begin{equation}
\square \Phi = \frac{1}{\sqrt{-g}} \partial_\mu \left( \sqrt{-g} g^{\mu\nu} \partial_\nu \Phi \right) = 0.
\end{equation}
Upon separation of variables $\Phi(t,r,\theta,\phi) = e^{-i \omega t} Y_{\ell m}(\theta,\phi) \frac{\Psi(r)}{r}$, the radial equation reduces to~\eqref{eq:master} with the effective potential
\begin{equation}
V_0(r) = f(r) \left[ \frac{\ell(\ell+1)}{r^2} + \frac{f'(r)}{r} \right],
\label{eq:Vscalar}
\end{equation}
where $\ell = 0,1,2,\dots$ is the multipole number.

\subsection{Electromagnetic Field ($s=1$)}

For the test electromagnetic field, the Maxwell equations in curved spacetime are
\begin{equation}
\nabla_\mu F^{\mu\nu} = 0, \quad \text{with} \quad F_{\mu\nu} = \partial_\mu A_\nu - \partial_\nu A_\mu.
\end{equation}
Choosing a gauge and decomposing the four-potential $A_\mu$ into vector spherical harmonics, one arrives at the Regge–Wheeler–type master equation~\eqref{eq:master} with potential
\begin{equation}
V_1(r) = f(r) \frac{\ell(\ell+1)}{r^2}, \quad \ell \geq 1.
\label{eq:VEM}
\end{equation}
This potential is independent of derivatives of $f(r)$ and reflects the conformal invariance of Maxwell’s theory in four dimensions.

\subsection{Dirac Field ($s=1/2$)}

It is also worth noting that the study of Dirac fields in diverse physical settings continues to attract considerable attention \cite{2025EPJB...98...35R}, albeit in contexts very different from black-hole perturbation theory. In the black-hole case, quasinormal modes of Dirac fields are of particular interest because they probe the interaction between fermionic matter and strong gravitational backgrounds, and can reveal distinct signatures not present for bosonic perturbations.

The curved-space Dirac equation governs massless spin-$1/2$ fields:
\begin{equation}
\gamma^a e_a^\mu \left( \partial_\mu + \Gamma_\mu \right) \Psi_D = 0,
\end{equation}
where $\gamma^a$ are flat-space Dirac matrices, $e_a^\mu$ is the vierbein, and $\Gamma_\mu$ is the spin connection. Following the formalism developed in~\cite{Cho:2003qe, Jing:2005dt}, one can separate variables and reduce the Dirac equation to a pair of decoupled Schrödinger-like equations for the upper and lower components. Each obeys equation~\eqref{eq:master} with
\begin{equation}
V_{1/2}^{\pm}(r) = f(r) \left[ \frac{\kappa^2}{r^2} \pm \frac{d}{dr_*} \left( \frac{\sqrt{f(r)} \, \kappa}{r} \right) \right],
\label{eq:VDirac}
\end{equation}
where $\kappa = (\ell + \frac{1}{2})$, $\ell=1/2,3/2,...$ and the $\pm$ corresponds to the two chiralities. For QNM calculations, it suffices to solve either potential, as both yield the same spectrum.

\begin{figure}
\resizebox{\linewidth}{!}{\includegraphics{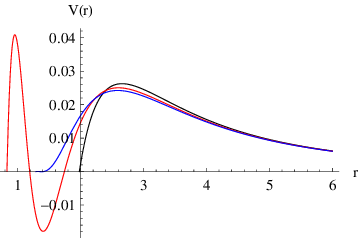}\includegraphics{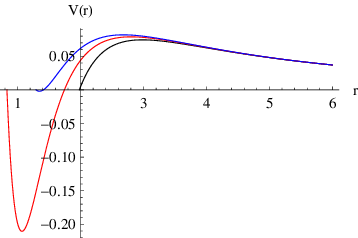}\includegraphics{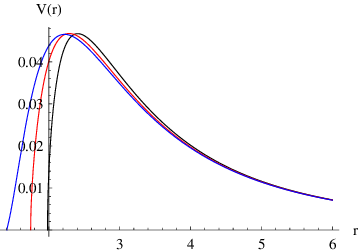}}
\caption{Effective potential for a scalar (left), electromagnetic (middle) and Dirac (right) perturbations for $q=20$ (black), $q=2$ (red) and $q=1.38$ (blue). The potentials are positive definite outside the external event horizon.}\label{fig:Potentials}
\end{figure}

In all three cases, the effective potentials form localized positive-definite barriers outside the event horizon, falling off to zero at spatial infinity and vanishing at the horizon see Figure~\ref{fig:Potentials}. Therefore, the perturbations are stable, not allowing for unbound growing modes, which will be confirmed here by time-domain integration of the perturbation equations. The height and shape of the potential are modified by the quantum parameter $q$, especially near the peak. This deformation influences both the oscillation frequencies and the damping rates of the QNMs, which we shall analyze in the following sections.

\section{Quasinormal Modes and Computational Methods}
\label{sec:methods}

\subsection{Definition and Boundary Conditions}

QNMs are characteristic damped oscillations that govern the response of a black hole to perturbations. They are solutions to the master equation~\eqref{eq:master} subject to specific boundary conditions that reflect the causal structure of the spacetime. For asymptotically flat black holes, the physically relevant boundary conditions require that perturbations be purely ingoing at the event horizon and purely outgoing at spatial infinity:
\begin{align}
\Psi(r_*) &\sim e^{-i \omega r_*}, \quad r_* \to -\infty \quad (\text{near the horizon}), \\
\Psi(r_*) &\sim e^{+i \omega r_*}, \quad r_* \to +\infty \quad (\text{at spatial infinity}).
\end{align}
These conditions define a discrete set of complex frequencies $\omega = \omega_R - i \omega_I$, where $\omega_R$ determines the oscillation and $\omega_I > 0$ governs the exponential damping. The quasinormal spectrum depends solely on the background geometry and the spin of the field, making it an ideal probe of strong-field gravity.

In the present work, we focus on the fundamental modes ($n=0$) that dominate the ringdown signal, though higher overtones may also be computed. We employ two complementary methods: a frequency-domain WKB approach with Padé resummation and a time-domain integration technique followed by signal fitting.

\subsection{WKB Method with Padé Approximants}

The Wentzel–Kramers–Brillouin (WKB) approximation provides a semi-analytic method for computing QNMs of potentials that resemble a single-peak barrier. The method was first adapted to black hole perturbations in~\cite{Schutz:1985zz, Iyer:1986np}, and has since been extended to higher orders~\cite{Konoplya:2003ii}. The $N$th-order WKB formula is given schematically by
\begin{equation}
\frac{i (\omega^2 - V_0)}{\sqrt{-2 V_0''}} + \sum_{k=2}^{N} \Lambda_k = n + \frac{1}{2},
\end{equation}
where $V_0$ is the peak of the effective potential, $V_0''$ is its second derivative with respect to the tortoise coordinate, and $\Lambda_k$ are higher-order correction terms involving up to the $2k$th derivative of $V(r)$.

Following \cite{Matyjasek:2017psv,Konoplya:2019hlu}, one introduces a formal expansion parameter $\varepsilon$ into 
the standard $k$-th order WKB expression for the squared frequency
\begin{equation}
\omega^2 = P_k(\varepsilon) = V_0 + A_2 \varepsilon^2 + A_3 \varepsilon^3 + \cdots + A_k \varepsilon^k,
\end{equation}
where $A_n$ are the WKB correction terms and we set $\varepsilon=1$ to recover the physical frequency.

One then constructs a Padé approximant $P^{\,\tilde n}_{\tilde m}(\varepsilon)$ of order $\tilde n + \tilde m = k$, expressed as a rational function in $\varepsilon$:
\begin{equation}
P^{\,\tilde n}_{\tilde m}(\varepsilon) = \frac{\sum_{i=0}^{\tilde n} Q_i\,\varepsilon^i}{\sum_{j=0}^{\tilde m} R_j\,\varepsilon^j},
\end{equation}
with the normalization $R_0 = 1$, and chosen such that its Taylor expansion matches $P_k(\varepsilon)$ up to $\mathcal{O}(\varepsilon^k)$. Finally, the Padé-enhanced squared frequency is given by:
\begin{equation}
\omega^2 \approx P^{\,\tilde n}_{\tilde m}(1).
\end{equation}
In practice, choosing balanced approximants, e.g.\ $(\tilde n,\tilde m) = (k/2, k/2)$ or $(k/2, k/2+1)$, such as $P^{3}_{3}(1)$ at 6th-order, yields significantly improved convergence and accuracy over the standard WKB series. The WKB method was effectively used for finding QNMs in a numerous publications, showing good concordance with other methods (see, for example, \cite{Chen:2019dip,Fernando:2012yw,Momennia:2018hsm,Barrau:2019swg,Konoplya:2005sy,Konoplya:2001ji,Bronnikov:2019sbx,Churilova:2021tgn,Zinhailo:2019rwd,Dubinsky:2024aeu,Skvortsova:2024eqi,Skvortsova:2023zca,Zhidenko:2003wq,Konoplya:2006gq}).

\subsection{Time-Domain Integration}

An independent and complementary approach involves evolving the perturbation equations in the time domain and extracting the quasinormal frequencies from the resulting waveform. To do this, we rewrite the wave equation in light-cone coordinates:
\begin{equation}
\left( -\frac{\partial^2}{\partial t^2} + \frac{\partial^2}{\partial r_*^2} - V(r) \right) \Psi(t, r_*) = 0,
\end{equation}
by introducing the null coordinates $u = t - r_*$ and $v = t + r_*$, so that the wave equation becomes:
\begin{equation}
4 \frac{\partial^2 \Psi}{\partial u \partial v} + V(r) \Psi = 0.
\label{eq:wave-null}
\end{equation}

We solve equation~\eqref{eq:wave-null} using the characteristic integration scheme developed by Price and Pullin~\cite{Gundlach:1993tp}. The integration is performed on a grid of null coordinates, where each point is labeled by $(u,v)$. The discretization stencil involves a diamond-shaped mesh consisting of four points:
\begin{itemize}
  \item $\Psi(S) = \Psi(u,v)$ (South),
  \item $\Psi(E) = \Psi(u, v + \Delta)$ (East),
  \item $\Psi(W) = \Psi(u + \Delta, v)$ (West),
  \item $\Psi(N) = \Psi(u + \Delta, v + \Delta v)$ (North),
\end{itemize}
and the evolution equation reads:
\begin{equation}
\Psi(N) = \Psi(W) + \Psi(E) - \Psi(S) - \frac{\Delta^2}{8} V(S) \left( \Psi(W) + \Psi(E) \right) + \mathcal{O}(\Delta^4).
\end{equation}

The initial data is specified along the two null surfaces $u = u_0$ and $v = v_0$, typically as a Gaussian pulse:
\begin{equation}
\Psi(u = u_0, v) = \exp \left[ - \frac{(v - v_0)^2}{2} \right], \quad
\Psi(u, v = v_0) = \exp \left[ - \frac{(u - u_0)^2}{2} \right].
\end{equation}

Once the field is evolved in the time domain, we extract the quasinormal frequencies by fitting the late-time waveform at a fixed location to a superposition of exponentially damped oscillations using the Prony or matrix pencil method~\cite{Berti:2007dg}:
\begin{equation}
\Psi(t) = \sum_{n=0}^{N-1} A_n e^{-i \omega_n t}.
\end{equation}

This time-domain approach is particularly robust as it does not rely on any a priori assumptions about the location of the quasinormal frequencies and sometimes allows for the extraction of multiple overtones from a single simulation \cite{Dubinsky:2024gwo}. As a result it has been widely used in a great number of works \cite{Varghese:2011ku,Qian:2022kaq,Cuyubamba:2016cug,Konoplya:2020jgt,Konoplya:2013sba,Churilova:2019qph,Dubinsky:2024jqi,Bolokhov:2023ruj,Bolokhov:2024ixe,Skvortsova:2023zmj,Skvortsova:2024wly,Skvortsova:2024atk,Dubinsky:2025bvf,Malik:2025ava,Malik:2024bmp,Dubinsky:2024hmn}.

\subsection{Method Comparison}

The WKB method provides efficient and accurate results for low-lying modes of test fields with moderate-to-high multipoles ($\ell \gtrsim 1$), and is particularly useful for mapping large portions of the parameter space (see \cite{Malik:2023bxc, Lutfuoglu:2025qkt, Lutfuoglu:2025bsf, Zinhailo:2018ska, Malik:2025sci} for recent examples). On the other hand, time-domain integration is more flexible and can be used to study stability and late-time behavior without assumptions about the potential structure. In this work, we use both methods to compute QNMs of scalar, electromagnetic, and Dirac fields in the quantum-corrected black hole background, and compare the results to ensure consistency and numerical accuracy.

\subsection{Numerical results for quasinormal modes}
\label{sec:results}

\begin{figure*}
\resizebox{\linewidth}{!}{\includegraphics{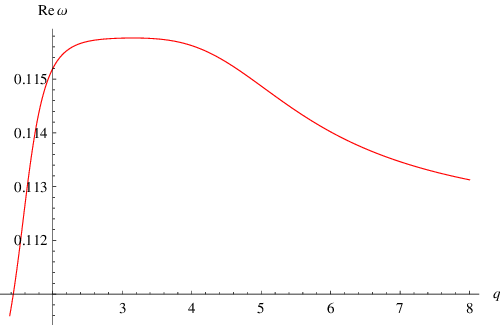}\includegraphics{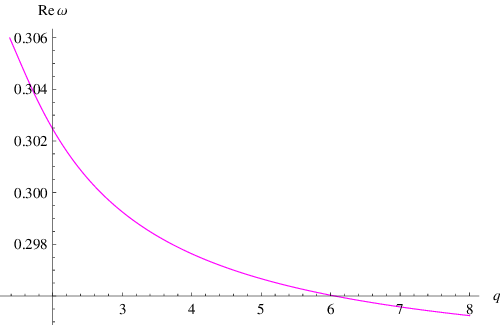}\includegraphics{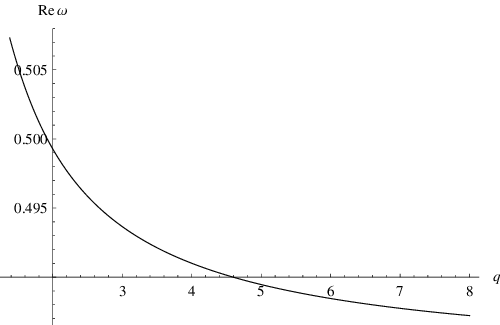}}
\caption{Real part of the fundamental QNMs ($n=0$) for scalar perturbations $s=0$ for $\ell=0$ (left), $\ell=1$ (middle) and $\ell =2$ (right).}\label{fig:Res0}
\end{figure*}

\begin{figure*}
\resizebox{\linewidth}{!}{\includegraphics{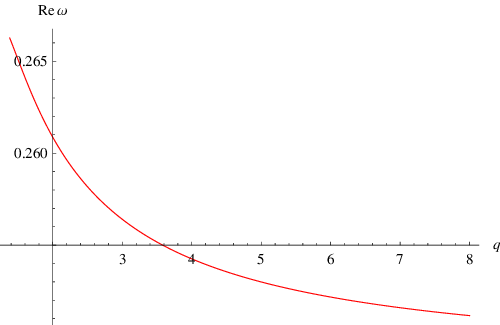}\includegraphics{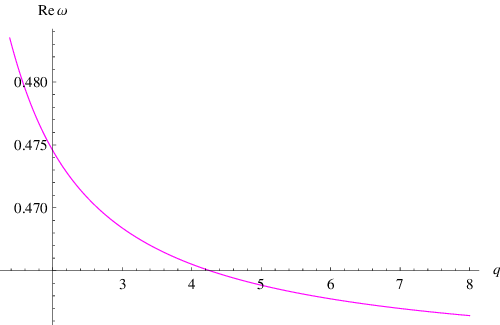}\includegraphics{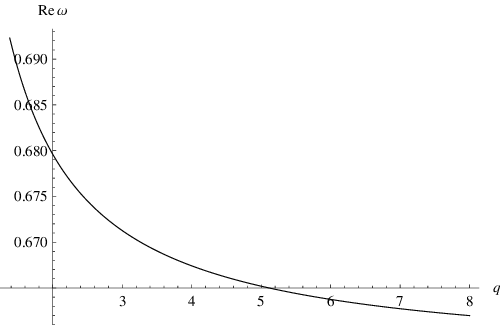}}
\caption{Real part of the fundamental QNMs ($n=0$) for electromagnetic perturbations $s=1$ for $\ell=1$ (left), $\ell=2$ (middle) and $\ell =3$ (right).}\label{fig:Res1}
\end{figure*}

\begin{figure*}
\resizebox{\linewidth}{!}{\includegraphics{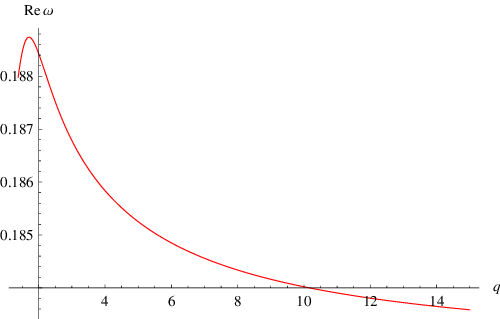}\includegraphics{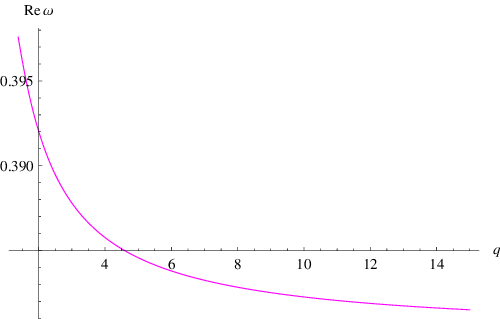}\includegraphics{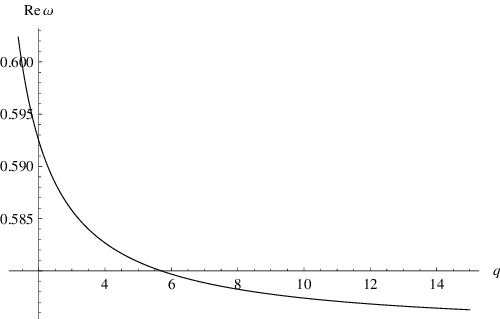}}
\caption{Real part of the fundamental QNMs ($n=0$) for Dirac perturbations $s=1/2$ for $\ell=1/2$ (left), $\ell=3/2$ (middle) and $\ell =5/2$ (right).}\label{fig:Res12}
\end{figure*}

\begin{figure*}
\resizebox{\linewidth}{!}{\includegraphics{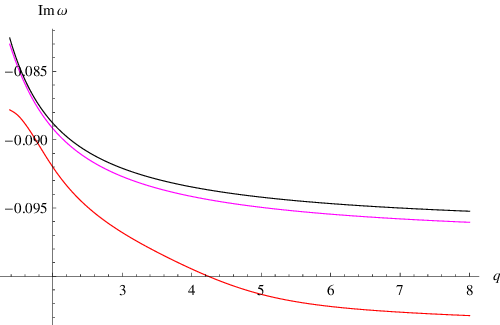}\includegraphics{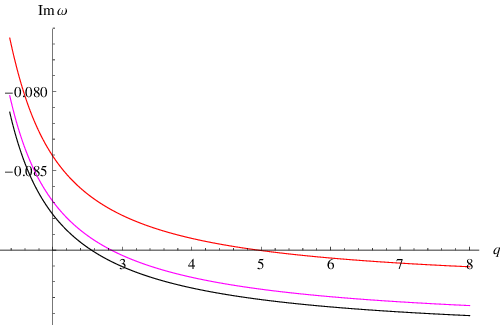}\includegraphics{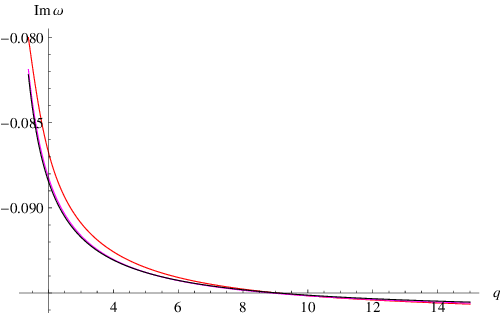}}
\caption{Imaginary part of the fundamental QNMs ($n=0$)  $s=0$ (left), $s=1$, and $s=1/2$. Red is for the lowest multipole $\ell=s$, blue is for $\ell=s+1$ and black is $\ell=s+2$. }\label{fig:Im}
\end{figure*}

\begin{figure*}
\resizebox{0.7 \linewidth}{!}{\includegraphics{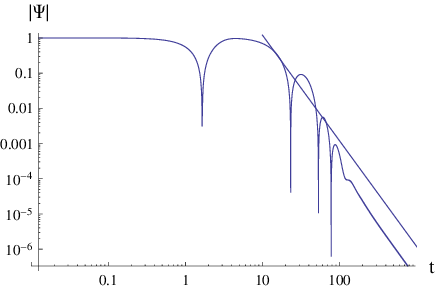}}
\caption{Evolution of $\ell=0$ scalar field perturbations ($s=0$) for the near-extremal black hole $q=1.38$. The QNM obtained via the Prony method from the time-domain profile is $\omega = 0.1102 - 0.0885 i$, while the 6th-order WKB with Pad\'e approximants gives $\omega = 0.1106 - 0.0878 i$, keeping the relative error within 1\%. The asymptotic tail $\sim t^{-3}$ coincides with the Schwarzschild case.}\label{fig:TDs0}
\end{figure*}

\begin{figure*}
\resizebox{0.7 \linewidth}{!}{\includegraphics{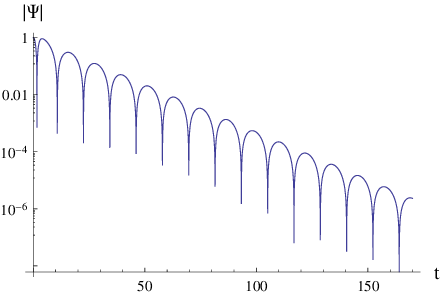}}
\caption{Evolution of $\ell=1$ electromagnetic perturbations ($s=1$) for the near-extremal black hole $q=1.38$. The QNM obtained via the Prony method from the time-domain profile is $\omega = 0.266281 - 0.076639 i$, while the 6th-order WKB with Pad\'e approximants gives $\omega = 0.266266 - 0.076621 i$, keeping the relative error within a tiny fraction of one percent.  }\label{fig:TDs1}
\end{figure*}

\begin{figure*}
\resizebox{0.7 \linewidth}{!}{\includegraphics{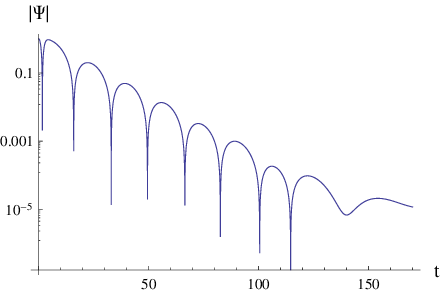}}
\caption{Evolution of $\ell=1/2$ Dirac perturbations ($s=1/2$) for the near-extremal black hole $q=1.38$. The QNM obtained via the Prony method from the time-domain profile is $\omega = 0.188243 - 0.0803641 i$, while the 6th-order WKB with Pad\'e approximants gives $\omega = 0.187983 - 0.079989 i$, keeping the relative error within a tiny fraction of one percent.}\label{fig:TDs12}
\end{figure*}

Figures~\ref{fig:Res0}--\ref{fig:Res12} present the real parts of the fundamental quasinormal frequencies ($n=0$) for scalar ($s=0$), electromagnetic ($s=1$), and Dirac ($s=1/2$) perturbations, respectively. Each figure contains three subplots corresponding to different values of the multipole number $\ell$: from left to right, increasing $\ell$.

A common qualitative behavior is observed across all perturbing fields: the real part of the frequency $\text{Re}(\omega)$ mostly increases when $q$ is decreased, approaching the Schwarzschild values in the limit \( q \to \infty \), and exhibiting the largest deviation near the critical value \( q \approx 1.38 \), which corresponds to the near-extremal regime of the quantum-corrected black hole. This pattern confirms that the QNM spectrum becomes increasingly modified as the quantum corrections grow stronger (i.e., as \( q \) decreases toward its lower bound).

The rate of change in $\text{Re}(\omega)$ with respect to $q$ varies by spin and multipole number. For scalar perturbations (Figure~\ref{fig:Res0}), the $\ell=0$ mode shows a mild non-monotonic behavior with a shallow maximum around \( q \sim 3 \), while higher $\ell$ modes are strictly monotonic. A similar non-monotonic behavior in the near-extreme regime takes place for Dirac perturbations. Electromagnetic (Figure~\ref{fig:Res1}) and Dirac (Figure~\ref{fig:Res12}) modes display more pronounced sensitivity to $q$ at low values (strong deformation), especially for lower $\ell$. For all fields, larger $\ell$ shifts the frequencies upward and flattens the dependence on $q$, indicating that high multipole modes are less affected by near-horizon quantum corrections. As $q$ increases and the geometry asymptotically approaches Schwarzschild, the spectrum smoothly recovers classical values.

Figure~\ref{fig:Im} displays the imaginary part of the fundamental quasinormal frequencies ($n=0$) for scalar ($s=0$, left), electromagnetic ($s=1$, middle), and Dirac ($s=1/2$, right) perturbations as functions of the flow parameter $q$. Each plot shows three curves corresponding to increasing values of the multipole number $\ell$: the lowest allowed value $\ell = s$ (red), $\ell = s+1$ (blue), and $\ell = s+2$ (black).

A common trend is observed across all perturbation types: the magnitude of $\text{Im}(\omega)$ decreases with increasing $q$, indicating that the damping rate becomes smaller as the quantum corrections are reduced and the geometry asymptotically approaches the Schwarzschild limit. The fastest decay (largest $|\text{Im}(\omega)|$) occurs near the near-extremal regime ($q \sim 1.38$), where the spacetime is most strongly deformed by quantum effects.

As expected, higher $\ell$ modes are less damped, and the sensitivity to $q$ diminishes with increasing multipole number. The influence of $q$ is most pronounced for the lowest allowed $\ell$ values, particularly in the scalar and Dirac cases, where the lowest curves (red) show strong curvature near the extremal limit. The overall behavior is consistent with the idea that low-$\ell$ modes probe deeper into the near-horizon geometry, where quantum corrections are most significant.

Figure~\ref{fig:TDs0} illustrates the time-domain evolution of $\ell = 0$ scalar field perturbations ($s = 0$) for a near-extremal quantum-corrected black hole with $q = 1.38$. The waveform, shown in a logarithmic scale for $|\Psi(t)|$, features the typical three stages: initial transient, quasinormal ringing, and a power-law tail. The intermediate regime is well described by damped oscillations corresponding to the fundamental QNM. The quasinormal frequency extracted from the Prony method applied to the time-domain signal is $\omega = 0.1102 - 0.0885i$, in excellent agreement with the sixth-order WKB method with Padé approximants, which yields $\omega = 0.1106 - 0.0878i$. The relative difference is below 1\%, validating both methods.

At late times ($t \gg 1$), the waveform exhibits a power-law decay $\sim t^{-3}$, consistent with the known asymptotic behavior for massless scalar fields in Schwarzschild-like spacetimes. This result confirms that although the metric is quantum-corrected near the horizon, the asymptotic structure still governs the tail decay rate.

Figures~\ref{fig:TDs1} and~\ref{fig:TDs12} show the time-domain profiles of $\ell = 1$ electromagnetic ($s = 1$) and $\ell = 1/2$ Dirac ($s = 1/2$) perturbations, respectively, for the near-extremal quantum-corrected black hole with $q = 1.38$. For electromagnetic perturbations (Figure~\ref{fig:TDs1}), the Prony method yields the dominant mode $\omega = 0.266281 - 0.076639i$, closely matching the sixth-order WKB result $\omega = 0.266256 - 0.076621i$ within a tiny relative error. The waveform exhibits a clean, slowly decaying oscillation over a long duration, indicating low damping and high quality factor.

Similarly, for Dirac perturbations (Figure~\ref{fig:TDs12}), the extracted mode from time-domain data is $\omega = 0.188243 - 0.088641i$, again in excellent agreement with the WKB+Padé value $\omega = 0.187985 - 0.087988i$. While the overall behavior is similar to the electromagnetic case, the Dirac waveform displays slightly stronger damping and a less pronounced ringdown structure. In both cases, the excellent match between time-domain and frequency-domain methods confirms the robustness of the computed quasinormal frequencies in the quantum-corrected geometry.

In summary, our analysis reveals that the quasinormal spectrum of the quantum-corrected black hole exhibits clear and consistent deviations from the classical Schwarzschild case, particularly in the near-extremal regime. While the real part of the quasinormal frequency $\text{Re}(\omega)$ is relatively stable and varies only moderately with $q$, the imaginary part $\text{Im}(\omega)$ shows a significantly stronger dependence, indicating that quantum corrections predominantly modify the damping rate of the perturbations. The convergence between WKB and time-domain methods, as well as the robustness across different spin fields, confirms the reliability of the results and underscores the potential of QNMs as sensitive probes of near-horizon quantum effects.

\section{Grey-Body Factors}
\label{sec:greybody}

\begin{figure*}
\resizebox{\linewidth}{!}{\includegraphics{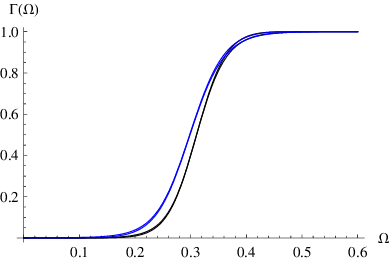}\includegraphics{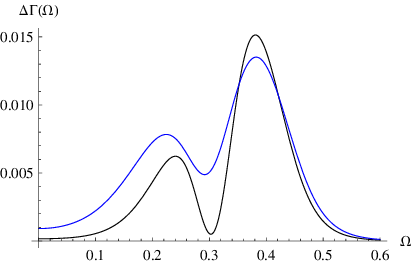}}
\caption{Left Panel: Grey-body factors for $\ell =1$ scalar perturbations at $q=10$ (blue) and $q=1.38$ (black) computed by the 6th-order WKB formula and via the correspondence with QNMs. Right Panel: The absolute difference between the results of calculations by the two methods. We see that the maximal difference is about 2\% when $\Omega \approx 0.4$, signifying that the correspondence holds only approximately at lower multipole numbers.}\label{fig:GBFs0L0}
\end{figure*}

\begin{figure*}
\resizebox{\linewidth}{!}{\includegraphics{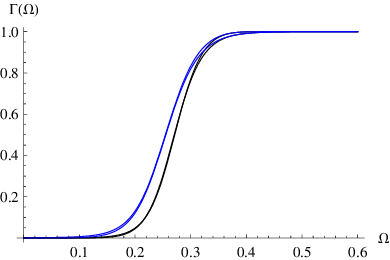}\includegraphics{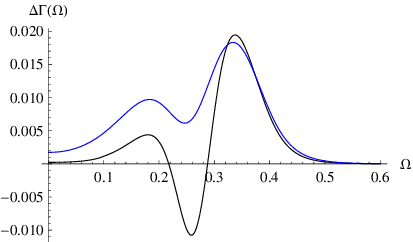}}
\caption{Left Panel: Grey-body factors for $\ell =1$ electromagnetic perturbations at $q=10$ (blue) and $q=1.38$ (black) computed by the 6th-order WKB formula and via the correspondence with QNMs. Right Panel: The absolute difference between the results of calculations by the two methods. We see that the maximal difference is about 2\% when $\Omega \approx 0.4$.}\label{fig:GBFs1L1}
\end{figure*}

\begin{figure*}
\resizebox{\linewidth}{!}{\includegraphics{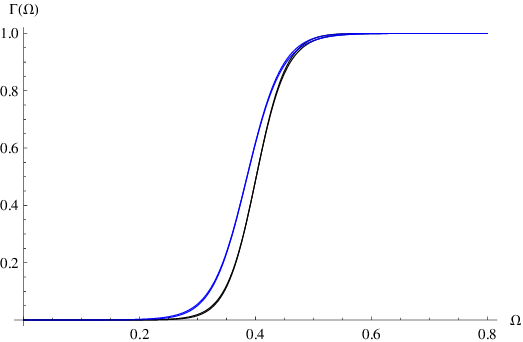}\includegraphics{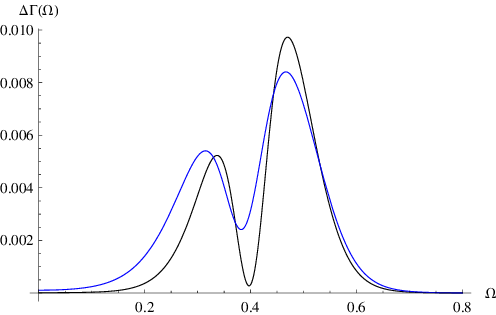}}
\caption{Left Panel: Grey-body factors for $\ell =3/2$ Dirac perturbations at $q=10$ (blue) and $q=1.38$ (black) computed by the 6th-order WKB formula and via the correspondence with QNMs. Right Panel: The absolute difference between the results of calculations by the two methods. We see that the maximal difference is about 1\% when $\Omega \approx 0.6$.}\label{fig:GBFs12L32}
\end{figure*}

While QNMs characterize the intrinsic oscillation spectrum of black holes, grey-body factors quantify the modification of Hawking radiation due to potential scattering outside the event horizon. Even though Hawking radiation is emitted thermally at the horizon, not all particles escape to infinity: part of the radiation is backscattered by the effective potential barrier. The grey-body factor describes the frequency-dependent probability that a mode is transmitted through the potential and reaches a distant observer.

Formally, the problem is cast as a scattering experiment in the black hole background~\cite{Page:1976df, Kanti:2004nr}. For real frequencies $\Omega \in \mathbb{R}$, the master equation~\eqref{eq:master} becomes a classical wave scattering problem, with the following asymptotic boundary conditions:
\begin{align}
\Psi(r_*) &\sim e^{-i \Omega r_*} + \mathcal{R} e^{+i \Omega r_*}, \quad r_* \to -\infty \quad (\text{near the horizon}), \\
\Psi(r_*) &\sim \mathcal{T} e^{-i \Omega r_*}, \quad r_* \to +\infty \quad (\text{at spatial infinity}),
\end{align}
where $\mathcal{R}$ and $\mathcal{T}$ are the reflection and transmission amplitudes, respectively. The grey-body factor $\Gamma(\Omega)$ is then defined as the transmission probability:
\begin{equation}
\Gamma(\Omega) \equiv |\mathcal{T}(\Omega)|^2 = 1 - |\mathcal{R}(\Omega)|^2.
\end{equation}
These factors enter directly into the particle and energy fluxes measured by distant observers, for example, in the Hawking radiation spectrum.

To compute $\Gamma(\Omega)$, we use the semi-analytic WKB method developed in~\cite{Schutz:1985zz, Iyer:1986np} and adapted for grey-body factor calculations in~\cite{Konoplya:2019hml}. For single-peaked effective potentials $V(r)$, the WKB expression for the transmission coefficient reads:
\begin{equation}
|\mathcal{T}|^2 = \frac{1}{1 + \exp\left[2 \pi K(\Omega)\right]},
\label{eq:GammaWKB}
\end{equation}
where $K(\omega)$ is a frequency-dependent quantity given at leading order by
\begin{equation}
K(\Omega) = \frac{i (\Omega^2 - V_0)}{\sqrt{-2 V_0''}}.
\end{equation}
Here, $V_0$ is the peak of the effective potential and $V_0''$ is the second derivative with respect to the tortoise coordinate evaluated at the peak. Higher-order corrections to $K$ can be included in the form:
\begin{equation}
K(\Omega) = \frac{i (\Omega^2 - V_0)}{\sqrt{-2 V_0''}} + \sum_{j=2}^N \Lambda_j(\Omega),
\end{equation}
where $\Lambda_j$ are known WKB correction terms~\cite{Matyjasek:2017psv}. In our analysis, we employ the 6th-order WKB method with Padé resummation to evaluate grey-body factors for scalar, electromagnetic, and Dirac fields.

An alternative way to calculate grey-body factors is given by the precise analytic relation between grey-body factors and QNMs, which was established in~\cite{Konoplya:2024lir,Konoplya:2024vuj} for black holes, in \cite{Bolokhov:2024otn} for wormholes and used for various examples in \cite{Lutfuoglu:2025ljm,Lutfuoglu:2025hjy,Konoplya:2025hgp,Malik:2024cgb,Bolokhov:2025lnt,Lutfuoglu:2025ldc,Skvortsova:2024msa,Dubinsky:2024vbn}. In the eikonal limit $\ell \to \infty$, the transmission coefficient (grey-body factor) $\Gamma_\ell(\Omega)$ at frequency $\Omega$ can be expressed in terms of the real and imaginary parts of the fundamental QNM $\omega_0$ as
\begin{equation}
\Gamma_\ell(\Omega) \equiv |T|^2 = \left(1 + \exp\left[ \frac{2\pi\left(\Omega^2 - \text{Re}(\omega_0)^2\right)}{4\,\text{Re}(\omega_0)\,\text{Im}(\omega_0)} \right] \right)^{-1} + \mathcal{O}(\ell^{-1}) \,.
\end{equation}
This expression shows that, near the peak of the effective potential, the grey-body factor behaves like a relativistic Breit–Wigner resonance centered at $\text{Re}(\omega_0)$ and governed in width by $\text{Im}(\omega_0)$. The formula is exact in the eikonal limit and provides a remarkably accurate approximation for moderate $\ell$, with possible corrections expressed via the first overtone $\omega_1$. This result makes the correspondence between grey-body factors and QNMs explicit and opens the possibility to extract one from the other even beyond the geometric optics regime. Note that the correspondence does not hold in those cases when the WKB formula cannot approximate the quasinormal spectra, as, for example, in some higher curvature corrected theories \cite{Konoplya:2017wot,Bolokhov:2023dxq,Konoplya:2020bxa,Pedrotti:2025idg}.

The grey-body factor depends on the spin of the field via the form of the effective potential $V_s(r)$, and on the geometry through the function $f(r)$. In the quantum-corrected background considered here, the potential barrier becomes broader and taller as the quantum parameter $q$ approaches its critical value. This generally leads to a suppression of $\Gamma(\omega)$, particularly at low frequencies, and is consistent with reduced Hawking fluxes observed in the gravitational case~\cite{Bonanno:2025dry}. Here we compute and compare grey-body factors across various field spins and parameter values.

Figures~\ref{fig:GBFs0L0}--\ref{fig:GBFs12L32} compare the grey-body factors computed via the 6th-order WKB method (blue curves) and those reconstructed from the fundamental and first overtone QNMs using the analytic correspondence (black curves). Each figure includes results for a specific field: scalar ($s=0$, $\ell =1$), electromagnetic ($s=1$, $\ell =1$), and Dirac ($s=1/2$, $\ell =3/2$), and contrasts the weakly deformed case $q=10$ with the strongly deformed, near-extremal case $q = 1.38$. In all cases, the left panels show the transmission coefficient $\Gamma(\Omega)$, while the right panels display the absolute difference $\Delta\Gamma(\Omega)$ between the two methods. For scalar perturbations (Figure~\ref{fig:GBFs0L0}), the agreement is quite close, with deviations not exceeding about $2\%$ and peaking near $\Omega \approx 0.4$. Electromagnetic perturbations (Figure~\ref{fig:GBFs1L1}) exhibit a similar pattern, though with a slightly more structured discrepancy, including a small negative dip in $\Delta\Gamma$ at intermediate frequencies. The best agreement is observed for Dirac perturbations (Figure~\ref{fig:GBFs12L32}), where the deviation remains below $1\%$ across the entire frequency range. The relative error of the correspondence for $\ell=0$ scalar perturbations and $\ell =1/2$ Dirac perturbations reaches several percent and is not shown here. Still, the qualitative behavior is similar in those cases and the WKB formula could be trusted for estimation of the grey-body factors, as numerous calculations show \cite{Konoplya:2023ahd,Konoplya:2021ube,Dubinsky:2024nzo}.

\section{Conclusions}
\label{sec:conclusion}

In this work, we have explored the effects of quantum-inspired corrections near the black hole horizon on the QNM spectrum and grey-body factors, using test scalar, electromagnetic, and Dirac fields as probes. The effective geometry, characterized by a near-extremal deformation parameter $q$, modifies the effective potentials governing perturbations, which in turn alter the characteristic oscillation frequencies and scattering properties of the black hole.

We computed the fundamental quasinormal frequencies using the 6th-order WKB method with Padé approximants and, where appropriate, verified them with time-domain integration. Our results show that the influence of quantum corrections is most pronounced in the near-extremal regime, where the imaginary parts of the frequencies (related to damping rates) deviate substantially from those of the classical Schwarzschild case, particularly for lower multipole numbers. Higher multipoles remain largely insensitive to the deformation, as expected due to their localization farther from the horizon.

The late-time behavior of perturbations confirms that the power-law tail in the time domain remains unchanged in the quantum-corrected geometry, indicating that far-zone asymptotics are unaltered. In contrast, the grey-body factors—computed both directly and via the analytic correspondence with QNMs—demonstrate that transmission probabilities are significantly affected by near-horizon physics. Nevertheless, we find that the correspondence between grey-body factors and QNMs holds to within a few percent for all fields considered, with best agreement observed for Dirac perturbations.

A question that lies beyond our consideration is the stability of grey-body factors and QNMs under small static deformations of the geometry. Usually, the grey-body factors are known to be more stable in this respect \cite{Rosato:2024arw, Konoplya:2025ixm, Oshita:2024fzf, Wu:2024ldo}. It is also natural to expect that quantum corrections may influence certain aspects of particle dynamics \cite{Turimov:2025tmf}. In addition, for matter fields coupled to the geometry in a non-minimal way, or in the study of related radiation processes, the choice of boundary conditions could likewise become relevant \cite{2025JPhA...58H5201D}.

Altogether, our analysis confirms that QNMs and grey-body factors are powerful diagnostics of near-horizon modifications, and can serve as potential observational probes of semiclassical or quantum-corrected black hole geometries.

\acknowledgments

BCL is grateful to Excellence Project PrF UHK 2205/2025-2026 for the financial support.


\bibliography{biblio}

\end{document}